\newcommand {\YBCO} [1] {${\rm YBa_2Cu_3O_{#1}}$}
\newcommand {\ket} [1] {|#1\rangle}
\newcommand {\bra} [1] {\langle#1|}
\documentstyle[aps,cite]{revtex}
\begin{document}
\draft
\title{Spin Susceptibility in Underdoped $\bf  YBa_2Cu_3O_{6+x}$}
\author{
H.F. Fong$^1$,
P. Bourges$^2$,
Y. Sidis$^2$,
L.P. Regnault$^3$,
J. Bossy$^4$,
A. Ivanov$^5$,
D.L. Milius$^6$,
I.A. Aksay$^6$,
B. Keimer$^{1,7}$}
\address{1 - Department of Physics, Princeton University, Princeton, NJ 
08544 USA}
\address{2 - Laboratoire L\'eon Brillouin, CEA-CNRS, CE Saclay, 91191 Gif 
sur Yvette, 
France}
\address{3 - CEA Grenoble, D\'epartement de Recherche Fondamentale sur
la mati\`ere Condens\'ee, 38054~Grenoble cedex~9, France}
\address{4 - CNRS-CRTBT, BP 166, 38042 Grenoble cedex~9, France}
\address{5 - Institut Laue-Langevin, 156X, 38042 Grenoble Cedex 9, France}
\address{6 - Department of Chemical Engineering, Princeton University,
Princeton, NJ 08544, USA}
\address{7 - Max-Planck-Institut f\"ur Festk\"orperforschung, 
70569 Stuttgart, Germany}

\twocolumn[\hsize\textwidth\columnwidth\hsize\csname@twocolumnfalse\endcsname
 
\maketitle

\begin{abstract}
We report a comprehensive polarized and unpolarized neutron scattering 
study of the evolution 
of the dynamical spin susceptibility with temperature and doping in three 
underdoped single 
crystals
of the \YBCO{6+x} high temperature superconductor:  \YBCO{6.5} ($\rm T_c = 
52$ K), 
\YBCO{6.7} ($\rm T_c = 67$ K), and \YBCO{6.85} ($\rm T_c = 87$ K). 
The spin susceptibility is determined in absolute units at 
excitation energies between 1 and 140 meV and temperatures between 1.5 and 
300 K. 
Polarization analysis is 
used extensively at low energies. 
Transitional matrix elements, including those between spin states, in a 
bilayer system such as  
\YBCO{6+x} can be generally classified into even and odd, according to the 
sign change under a 
symmetry operation that exchanges the layers, and both even and odd 
excitations are detected in
\YBCO{6.5} and \YBCO{6.7}. While the even spin excitations show a true gap 
which depends on doping, the odd spectrum is characterized by a weakly 
doping-dependent 
pseudogap. Both even and odd components are substantially enhanced upon 
lowering the 
temperature from 300 K. The even excitations evolve smoothly through the 
superconducting 
transition temperature $\rm T_c$, but the odd excitations develop a true 
gap below $\rm T_c$. 
At the same time, the odd spin susceptibility is sharply enhanced below 
$\rm T_c$ around an 
energy  that increases with doping. This anomaly in the magnetic spectrum 
is 
closely related to the magnetic resonance peak that appears at 40 meV in 
the superconducting 
state of the optimally doped 
compound  ($\rm T_c = 93$ K). From these data we extract the energy and 
the energy-integrated 
spectral weight of the resonance peak in absolute units as a function of 
doping level. Theoretical 
implications of these measurements are discussed, and a critique of recent 
attempts to relate the 
spin excitations to the thermodynamics of high temperature superconductors 
is given.
\end{abstract}

]

\section{Introduction}
The importance of electronic correlations in the copper oxide high 
temperature superconductors 
is now generally recognized. Some of the strongest evidence has been 
provided by neutron 
scattering measurements of the spin excitations in several families of 
cuprates: $\rm La_{2-
x}Sr_xCuO_4$ \cite{kastner98}, \YBCO{6+x} 
\cite{rossat93,revue-cargese,revue-lpr,rossat91,mook93,
tranquada92,sternlieb94,fong95,bourges96,fong96,dai96,
fong97,bourges97_euro,bourges97_prb,keimer97,keimer98,miami,dai97,mook98,fong99}, 
and recently 
also $\rm Bi_2 Sr_2 Ca Cu_2 O_{8+\delta}$ \cite{fong99_nature}.  
While in an uncorrelated metal the spin excitation spectrum takes the form 
of a broad continuum 
extending up to energies comparable to the Fermi energy, the spin 
excitations in the metallic 
copper oxides in many ways resemble the antiferromagnetic magnons in their 
insulating parent 
compounds. This is illustrated by the mere fact that spin excitations 
cannot be observed at all in the 
metallic materials by neutron scattering: the spectral weight of continuum 
excitations in an 
uncorrelated metal is more than an order of magnitude below the 
sensitivity limit of current 
neutron instrumentation. In this article we continue our program of 
putting this qualitative 
observation on a quantitative footing by converting the neutron cross 
section to the absolute spin 
susceptibility.

The optimally doped superconductors \YBCO{7} and $\rm Bi_2 Sr_2 Ca Cu_2 
O_{8+\delta}$ 
with transition temperatures in the 90 K range at first sight appear 
compatible with a weak 
correlation picture as the normal-state spin susceptibility is indeed too 
weak to be measurable 
with the instrumentation currently available 
\cite{bourges96,fong96,fong99_nature,miami}. However, 
below the superconducting transition 
temperature both compounds develop a remarkably sharp ``resonance peak'' 
at an energy of 40 
meV whose spectral weight is comparable to that of antiferromagnetic 
magnons at the same 
energy \cite{rossat91,mook93,tranquada92,sternlieb94,fong95,bourges96}. 
This discovery, 
which was followed up by related observations
in underdoped \YBCO{6+x} \cite{dai96,fong97,bourges97_euro} 
has generated much theoretical work, most of which is based on 
strong-correlation models 
\cite{mazin95,millis96,bulut96,blumberg95,liu95,stemmann94,lavagna93,onufr
ieva95,
onufrieva95_1,onufrieva99,barzykin95,morr98,assaad98,zhang97,demler95,deml
er98_prb,weng98,
moriya99,brinckmann99}.
Kinetic energy effects such as band structure singularities 
\cite{bulut96,blumberg95,abrikosov98} or the 
manifestations of interlayer pair tunneling along the c-axis \cite{lin97} 
have, however, also been 
implicated. The 
resonance peak has not been observed in conventional superconductors, but 
some recent data on 
heavy fermion compounds appear to have a similar signature \cite{fermion}. 
Here we give a 
comprehensive account of this unusual 
phenomenon that provides important clues to the mechanism of high 
temperature 
superconductivity. We fully describe the evolution of its energy and 
spectral weight with hole 
concentration and elucidate the relationship of the resonance peak to the 
normal state 
excitations in the underdoped regime. 

Our new experiments are generally consistent with previous investigations 
of the spin dynamics in the underdoped regime
\cite{rossat93,revue-cargese,revue-lpr,rossat91,tranquada92,sternlieb94}, 
but due to the combination of large single crystals and advanced neutron 
instrumentation the energy range could be extended above 100 meV, 
comparable to the 
nearest-neighbor antiferromagnetic superexchange energy $J$ in the 
insulator, and the statistics 
could be 
substantially improved. Due to the extended energy range, we were able to 
probe not only the low 
energy sector of the spin excitation spectrum which evolves out of in-
phase precession modes of 
spins in directly adjacent layers of the antiferromagnetic insulator 
\YBCO{6} (acoustic 
magnons, Ref. \cite{rossat93,tranquada89}), but also the higher 
energy sector evolving out of  antiphase (optical) 
magnons \cite{reznik96}. In the 
metallic state we prefer to characterize these excitations by their 
symmetry (odd or even, 
respectively) under exchange of the two layers. Observation of excitations 
in both sectors is 
important because 
it throws light on interlayer interactions within a bilayer both in the 
normal and in the 
superconducting states, an issue that has emerged as an important theme in 
the theory of high 
temperature superconductivity.

The article is organized as follows. In Section \ref{sec:def} we introduce 
our notation and the 
absolute scale 
which underlies much of the presentation. Technical details of the sample 
characterization and 
the neutron 
experiments are given in Sections \ref{sec:sample} and \ref{sec:expt}, 
respectively. 
Section \ref{sec:results} gives a comprehensive account of the results on 
even and odd excitations in the normal state, and the spin gap and 
magnetic resonance peak in the 
superconducting state. The data are discussed in Section 
\ref{sec:discussion} in the light of recent 
model 
calculations. Preliminary accounts of this work have been given in 
\cite{fong97,bourges97_prb,keimer97,keimer98,miami}.

\section{Definitions and Notation}
\label{sec:def}
\subsection{Magnetic Susceptibilities and Absolute Units}

Throughout this paper, wave vectors are expressed in reciprocal lattice 
units (r.l.u.), that is, 
${\bf Q} = H {\bf a^*} +  K {\bf b^*} + L {\bf c^*}$ with ${\bf a^*} \sim 
{\bf b^*} \sim 1.63 
{\rm \AA}^{-1}$ 
and ${\bf c^*} \sim 0.57 {\rm \AA}^{-1}$.
In these units, the antiferromagnetic zone center commonly referred to as 
${\bf Q}_0 = (\pi,\pi)$,  
is at
${\bf Q}_0 = (\frac{1}{2}, \frac{1}{2})$.
The two-dimensional reduced wave vector, modulo ${\bf Q}_0$, is denoted by 
{\bf q}.

Magnetic neutron scattering probes the
imaginary part of the magnetic susceptibility tensor \cite{lovesey},
\begin{eqnarray}
\frac{d^2\sigma}{d\Omega\,dE} &=& r_0^2 \frac{k_f}{k_i} f^2({\bf Q})
\exp(-2W({\bf Q})) \frac{N}{\pi(g\mu_B)^2} [1+n(\omega)] 
\nonumber\\
&& \sum_{\alpha\beta} 
(\delta_{\alpha\beta}- \frac{Q_\alpha Q_\beta}{Q^2})
\chi''_{\alpha\beta}({\bf Q},\omega)   \label{crosssection}
\label{d2sigma}
\end{eqnarray}
where $r_0$ is the classical electron radius, ${\bf k}_i$ and ${\bf k}_f$ 
are the wave vectors of
the incident and scattered neutron with ${\bf Q}={\bf k}_f - {\bf k}_i$,
$\alpha, \beta$ are the spatial components of the tensor, $f({\bf Q})$ is 
the atomic form factor of the 
orbitals that 
contribute to the
inelastic process, $W({\bf Q})$ is the Debye-Waller exponent, and $N$ is 
the number of spins in 
the system. 
The susceptibility is related to the spin-spin correlation function in the 
form:
\begin{eqnarray}
{\rm S}_{\alpha\beta}({\bf Q},\omega) &=& \frac{1}{2\pi\hbar N} \sum_{ij} 
\int dt\,e^{i {\bf Q} \cdot ({\bf R}_j - {\bf R}_i) -i\omega t}
\langle \hat S^\alpha_i \hat S^\beta_j(t) \rangle  \nonumber\\
&=& \frac{1+n(\omega)}{\pi(g\mu_B)^2} \chi''_{\alpha\beta}({\bf Q},\omega)
\label{eq-corrchi}
\end{eqnarray}
where $S_i$ is the usual spin operator at site ${\bf R}_i$.
It will often be fruitful to refer to this quantity because there is a sum 
rule
for it (see Eq. \ref{eq-sumrule} below).

When there is no magnetic long range order in the system and 
no preferred orientation, the summation in Eq. \ref{crosssection} 
must be rotationally invariant, that is,
$\sum_{\alpha\beta}  (\delta_{\alpha\beta}- \frac{Q_\alpha Q_\beta}{Q^2})
\chi''_{\alpha\beta}({\bf q},\omega)=2 {\rm Tr}(\chi''_{\alpha\beta})/3$.
Therefore, using the isotropic susceptibility
$\chi'' \equiv {\rm Tr}(\chi''_{\alpha\beta})/3$,  Eq. \ref{d2sigma}
becomes for a magnetically disordered system
\begin{equation}
\frac{d^2\sigma}{d\Omega\,dE} = 2 r_0^2 \frac{k_f}{k_i} f^2({\bf Q})
\exp(-2W({\bf Q})) \frac{1+n(\omega)}{\pi(g\mu_B)^2}
\; \chi''.
\end{equation}

We had already used this definition previously  \cite{bourges97_prb}.
However, in an earlier paper, some of us \cite{fong96} used 
a slightly different convention for the dynamical susceptibility:
\begin{equation}
\frac{d^2\sigma}{d\Omega\,dE} = r_0^2 \frac{k_f}{k_i} f^2({\bf Q})
\exp(-2W({\bf Q})) (1+n(\omega)) \frac{\hat{\chi}''}{3},
\end{equation}
so that $\chi'' = ( 2\pi\mu_B^2/3)\, \hat{\chi}''$. 
The current convention agrees with the work of Hayden {\it et al}. \cite{hayden96} 
in the metallic state of  $\rm La_{2-x} Sr_x CuO_4$ as well as 
with their recent work in \YBCO{6+x} \cite{dai99}. Throughout this paper, 
$\chi''({\bf Q},\omega)$ is given in $\mu_B^2/{\rm eV}/[{\rm unit\ 
formula}].$

The signal detected in a neutron scattering experiment is proportional to 
the intrinsic $\chi''$ 
convoluted with the instrumental resolution function. We extract 
$\chi''({\bf Q},\omega)$ from 
the data 
using a model of the resolution function and convert the result to 
absolute units through a 
calibration against selected phonons (see Appendix). Depending on the 
structure of the magnetic 
spectrum in momentum space, it is often convenient to further process the 
data by integrating 
$\chi''$ over momentum or energy. We henceforth often use the spectral 
weight, here defined as 
$\int d(\hbar \omega) \;
\chi''({\bf Q}_0,\omega)$, as well as the  Brillouin zone averaged 
susceptibility $\chi''_{\rm 
2D}(\omega)$, 
also known as local susceptibility, defined as
\begin{equation}
\chi''_{\rm 2D}(\omega) \equiv \frac{\int d^2q \; \chi''({\bf 
q},\omega)}{\int d^2q}.
\label{chi2D}
\end{equation}

The latter quantity is often used for systems with dispersing excitations. 
In particular, for a 
two-dimensional (2D) Heisenberg antiferromagnet, we have at low energies 
(where
the magnon dispersion relation is approximately linear in wave vector):
\begin{equation}
\chi''_{\rm 2D}(\omega) \sim 4SZ_\chi \mu_B^2/J \sim 10\; \mu_B^2/{\rm 
eV},
\label{chiAF}
\end{equation}
independent of energy. We here used the theoretical value of 
quantum corrections for $S={1\over2}$, $Z_\chi = 0.51$ \cite{igarashi},
although neutron scattering experiments in several undoped cuprates 
\cite{hayden96,jin} 
indicate a systematic reduction of $\sim$ 30 \% of the spin-wave spectral weight
as compared with the theoretical estimate (possibly due to covalency effects). Nevertheless, 
Eq. \ref{chiAF} provides a useful benchmark for the ``oscillator 
strength'' of normal state magnetic excitations in the metallic cuprates. 
The magnetic resonance 
peak in the superconducting state, on the other hand, is a non-dispersive 
excitation centered 
around a single point $\bf{Q_0}$ in momentum space. As it is very narrow 
in energy, the instrumental resolution function carries out a partial
energy-integration so that its spectral 
weight is best expressed in terms of $\int d(\hbar \omega) \; \chi''({\bf 
Q_0},\omega)$.

\subsection{Even and Odd Excitations}
\YBCO{6+x} is a bilayer system, that is, two closely spaced copper oxide 
layers are separated by 
a much 
larger distance. Information about the magnetic coupling between the 
layers has been obtained 
from studies of the spin wave dispersions in insulating \YBCO{6.2}. The 
superexchange 
coupling between 
adjacent bilayers is more than four orders of magnitude smaller than the 
primary energy scale, $J 
\sim 100$ meV \cite{rossat93,tranquada89}. 
In metallic \YBCO{6+x}, where the 
in-plane correlation length is never larger than a few lattice spacings, 
the bilayers are therefore 
expected to be magnetically decoupled from each other, so that the neutron 
scattering signal is an 
incoherent superposition of the signals arising from individual bilayers. 
This is in accord with the 
experimental situation.

The intra-bilayer coupling $ J_\perp \sim 10$ meV \cite{reznik96}, on the other hand, 
is only an order of magnitude below the intralayer superexchange, 
and it has long been known that even in the metallic regime the 
two layers within a bilayer remain strongly 
coupled\cite{rossat93,revue-cargese,revue-lpr,tranquada92}. 
One focus of the present study is 
to further elucidate this magnetic interlayer coupling.

To this end, we consider an eigenstate $\ket{n}$ of the in-plane crystal 
momentum centered on 
the layer $n$. We can then define symmetric and antisymmetric combinations 
of states centered 
on the two layers within a bilayer:
\begin{eqnarray}
\ket{s} &=& ( \ket{1} + \ket{2} )/\sqrt{2} \nonumber\\
\ket{a} &=& ( \ket{1} - \ket{2} )/\sqrt{2} 
\end{eqnarray}
In this new basis, elementary excitations 
can be characterized by transitions between states of 
the same or opposite symmetry (even or odd excitations,
respectively). The contribution of these two types of transition between 
spin states to the neutron 
scattering signal can be varied deliberately by adjusting the 
momentum transfer
component $Q_z$ perpendicular to the copper oxide layers. 
In particular, with layer states $\ket{1}$ and $\ket{2}$,
\begin{eqnarray}
\hat{z} \ket{1} &=& d/2 \ket{1} \nonumber \\
\hat{z} \ket{2} &=& -d/2 \ket{2} \nonumber \\
\bra{a,s} e^{iQ_z \hat{z}} \ket{a,s} &=& \cos(Q_z d/2) {\rm\ 
(even)}\nonumber \\
\bra{s,a} e^{iQ_z \hat{z}} \ket{a,s} &=& i \sin(Q_z d/2) {\rm\ (odd)}
\end{eqnarray}
where $d$ is the intrabilayer distance. Therefore, the intensities of 
the odd and even excitations, regardless of their in-plane wave vector and 
energy
dependence, follow the $\sin^2(Q_z d/2)$ and $\cos^2(Q_z d/2)$ 
modulation, respectively, and we can write, following Eq. \ref{eq-corrchi},
\begin{eqnarray}
\chi''({\bf Q},\omega) &=& \chi''_{\rm odd}({\bf q},\omega) 
\sin^2(Q_z d/2) \nonumber\\
&&+ \chi''_{\rm even}({\bf q},\omega) \cos^2(Q_z d/2).
\end{eqnarray}
In the antiferromagnetic insulator,  $\chi''_{\rm odd}$
and $\chi''_{\rm even}$ correspond to the acoustic 
and optical spin-waves, respectively \cite{reznik96}. 
The local and the energy-integrated 
susceptibilities as defined in the previous subsection
can likewise be decomposed into their even and odd components. 

\section{Crystal Synthesis and Characterization}
\label{sec:sample}
Our samples are three single crystals of volume ${\rm \sim 1-3\; cm^3}$.
A precursor powder was made by combusting a stoichiometrically mixed
aqueous solution of nitrates with sucrose. The powder was then 
calcined in a fully ventilated environment at 940$^\circ$C for 6 hours to 
form YBCO powders.  
Crystals were grown using top seeded sintering techniques \cite{gautier94}. 
After fully oxygenating the resulting crystal by annealing in a pure 
oxygen environment, oxygen 
was systematically removed by progressive annealing under argon flow until
the desired fraction of oxygen remained. They were then sealed in an 
evacuated quartz tube and 
annealed for two weeks to ensure homogeneity. 
We estimate the oxygen content by the weight gain/loss of the sample 
during the anneal, in 
conjunction with 
comparing to the $\rm T_c$ calibration of Cava {\it et al.} \cite{cava}. 
Several factors contribute to 
a rather large 
systematic error ($\Delta x \sim \pm 0.05$) in the estimate of the 
absolute oxygen content. For 
instance, a 
significant fraction of a foreign phase ($\rm Y_2 Ba Cu O_5$, the so-called
green-phase) embedded in the single crystal matrix makes it 
difficult to translate the weight change of the crystal into a change in 
stoichiometry. 
Three crystals were used: \YBCO{6.5} with ${\rm T_c \sim 52}$ K and mass 
23.3 g,
\YBCO{6.7} with ${\rm T_c \sim 67}$ K and mass 14.8 g, and \YBCO{6.85} 
with ${\rm T_c 
\sim 87}$ K and mass 9.5 g.

The superconducting properties of these samples were characterized by SQUID 
magnetometry. 
The sharpness of the diamagnetic transition is usually a good indicator, 
but by no means a 
guarantee, of the uniformity of the oxygen distribution in a large crystal. 
A surface sheet of high 
oxygen content material with a high $\rm T_c$ may in principle mask more 
poorly 
superconducting material in the interior. At the expense of reducing the 
sample volume, the 
SQUID measurements 
on the \YBCO{6.5} sample were therefore performed on a piece cut from the 
interior of 
the sample. Details of these measurements have been given elsewhere 
\cite{fong97}. 
On the \YBCO{6.7} sample, we have, in addition, employed a nondestructive neutron 
depolarization 
technique that is sensitive to the distribution of superconducting 
transition temperatures 
throughout the bulk of the sample. This method is described in Section \ref{sec:expt}
below.

\section{Experimental details}
\label{sec:expt}
The crystals were aligned such that wave vector transfers of the form 
$(H,H,L)$ or 
$(3H,H,L)$ were accessible and loaded into a cryostat.
The initial experiments were carried out at Brookhaven National Laboratory 
(BNL), on the
H4M, H7 and H8 thermal triple axis spectrometers of the High Flux Beam 
Reactor.
The bulk of the work was then carried out on the 2T spectrometer at the 
Laboratoire L\'eon 
Brillouin (LLB) for the \YBCO{6.7} sample, and on the hot-neutron IN1 and 
thermal-neutron 
IN8 spectrometers at the Institut Laue-Langevin (ILL) for the \YBCO{6.5} 
sample. Additional 
polarized-beam experiments were done on IN20 at the ILL.
At BNL, we used pyrolytic graphite (PG) (002) monochromators and analysers 
with a PG filter 
in the scattered beam and neutrons of 30.5 meV fixed final energy. 
Collimations were set at 
40-40-80-80. At the ILL, the incident beam was monochromated by a 
vertically focused Cu(111) 
crystal on IN8, and Cu(200) or Cu(220) with vertical focusing were used as 
monochromators 
on IN1. Vertically and horizontally focusing PG(002) was used as analyser. 
At LLB, we used a 
vertically curved Cu(111) monochromator and a vertically and horizontally 
focusing PG(002) 
analyser. Heusler(111) crystals were used in the polarization work at 
IN20. No beam collimations were used at the ILL and the LLB in order to maximize 
the benefits of 
focusing. A PG filter was placed behind the sample on 2T and IN8, and the
final energy was fixed at 14.7, 30.5, or 35 meV depending on the range of 
excitation energies 
covered. An Erbium filter was used on IN1, and the final energy was 
fixed at 62.6 meV. The energy and momentum resolutions varied depending on the
instrumental configuration used. Typical values for the energy resolution
are 1 meV (full width at half maximum, FWHM) at energy transfer 5 meV,
5 meV at 50 meV, and 12 meV at 100 meV. The vertical resolution
was typically $0.35 \AA^{-1}$ FWHM. Full four-dimensional resolution calculations
were used to extract the instrinsic magnetic neutron cross section from
the raw data.

The Heusler monochromator used in the polarized beam experiments \cite{moon,fong96} reflects 
only neutrons of a specific 
(vertical) spin polarization direction. Before impinging on the sample, the beam
polarization is maintained (vertical field, VF) or rotated by $\rm 90^\circ$
(horizontal field, HF) by homogeneous guide fields. After scattering from the sample, the 
beam polarization is again maintained or rotated back by $\rm 90^\circ$, 
respectively. The beam then traverses a flipper (a set of coils capable of
flipping the neutron spin polarization by $\rm 180^\circ$), and the
final beam polarization is analysed by a Heusler crystal which Bragg-reflects
only neutrons whose polarization direction is the same as that of the original
beam (after the monochromator). Because of limitations of the apparatus the beam polarization is always
incomplete and is usually parametrized as (FR-1)/(FR+1), where
FR is the ``flipping ratio''. When the flipper is on, the spin-flip (SF) 
cross section is measured, superposed by a polarization
``leakage'' contribution from non-spin-flip (NSF) scattering events 
(mostly phonon scattering), a 
contribution from nuclear spin incoherent scattering (NSI), and an 
extrinsic background (B). Because of polarization terms in the coherent
magnetic scattering cross section \cite{moon} only half of the magnetic contribution (M) is 
measured for vertical guide field, whereas for HF the full contribution is
measured. When the flipper is on and the flipping ratio is not too small, one 
obtains \cite{moon}

\begin{eqnarray}
{\rm I_{\rm HF}} & = & {\rm M + \frac{2}{3} NSI + \frac{NSF}{FR} + B} 
\nonumber \\
{\rm I_{\rm VF}} & = & {\rm \frac{1}{2} M + \frac{2}{3} NSI +
\frac{NSF}{FR} + B} 
\label{hfvf}
\end{eqnarray}

The standard method of extracting the magnetic contribution to the cross
section is to subtract I$_{\rm VF}$ from I$_{\rm HF}$, which yields M/2. 

We used this setup on IN20 in order to determine the superconducting transition
temperature of one of our large single crystals (\YBCO{6.7}) nondestructively and {\it in situ}. 
The technique relies on the fact that a 
spin-polarized neutron 
beam is depolarized when 
transiting a region containing a magnetic field whose direction varies on 
a short length scale, such 
that the neutron spins do not follow the change in field direction 
adiabatically. Depolarization is negligible only if the parameter 
$\eta = \gamma B / v \frac{d\theta}{dx} >> 1$, where $\gamma = 1.8 \times 10^{8} \, {\rm T}^{-1} {\rm sec}^{-1}$ 
is the neutron gyromagnetic ratio,
$B \sim 10$ G the applied field, $v \sim 2000$ m/sec the neutron velocity, and 
$\frac{d\theta}{dx}$ the directional variation of $B$ \cite{zhuchenko97}. 
The spectrometer is set for 
a nuclear Bragg reflection. The field $B$ is applied at the sample position in the same 
direction as the guide field 
before and after the sample, so that the beam polarization is maintained 
over the neutron flight 
path and the full nuclear Bragg intensity is detected. This remains 
unchanged as the sample is 
field-cooled through the superconducting transition. Because of the porous microstructure
of our samples due to inclusions of a second phase, $\rm Y_2 Ba CuO_5$, 
magnetic flux penetrates even at low fields and gets trapped by microstructural
defects such as twin boundaries and the $\rm Y_2 Ba CuO_5$ inclusions. This flux 
has the 
same direction as the guide field and therefore does not depolarize the 
beam. The field is subsequently turned by 90 degrees at low temperature. 
Because of flux pinning, the vortices do not follow this change 
of field orientation, so that the net field at the sample position 
is no longer parallel to the guide field. Significant beam depolarization 
results because $\frac{d\theta}{dx} \sim \frac{\pi/2}{1 \, {\rm cm}}$, where 1 cm is a typical sample dimension,
and hence $\eta < 1$. The measured 
intensity of the nuclear Bragg reflection is therefore reduced (Fig. \ref{fig-6.7Tc}). Heating the sample in 
this state and monitoring the Bragg intensity provides a $\rm T_c$-curve 
characteristic of the 
entire bulk of the sample. The small transition width shown in Fig. 
\ref{fig-6.7Tc} attests to the 
high quality of the \YBCO{6.7} sample.

While neutron scattering with polarization analysis is the most powerful 
tool to
study magnetic excitations, current instrumentation restricts its use to 
rather low energy transfers
(below $\sim$ 50 meV). Moreover, the neutron flux is substantially lower 
than in a standard 
unpolarized-beam 
experiment. For an extensive survey of the magnetic spectrum, and for 
higher excitation energies,
inelastic neutron scattering experiments without polarization analysis are 
the only option. In 
unpolarized-beam 
experiments, the dynamic structure factor of phonons can exhibit various 
features in momentum 
space that 
can be mistaken for magnetic fluctuations at the worst, or complicate
background subtraction at the very least. In order to arrive at a
reasonable description of the phonon background, we numerically
modeled the lattice dynamics of \YBCO{6+x} using a simple harmonic 
interaction model. Once 
the phonon dispersion relations and eigenvectors are known as a function 
of wave vector, 
their full cross section can be calculated in absolute units. This has the 
added benefit of providing 
a standard against which the magnetic cross section can be calibrated.
In previous publications, we have demonstrated the success of this model 
in predicting the cross 
section of a particularly important phonon at 42.5 meV 
\cite{fong95,fong96}. 
We have since augmented our calibration procedure by including several 
acoustic phonons, and 
we have further improved the parameters used in the simulation. Details 
are given in
the Appendix.

We do not attempt to fit the whole phonon spectrum with our
somewhat simplified model of the lattice dynamics which is 
still not completely understood  \cite{karlsruhe}. Nevertheless,
the simulation has given us a useful guide to extract the magnetic
signal in some situations. As an example, Fig. \ref{fig-sim} shows a 
``worst-case'' scan at 55 
meV, an energy 
range in which the phonon cross section is particularly large in the 
region of momentum space 
where the magnetic cross section is peaked. This excitation energy is too 
large for polarized-beam 
experiments. The figure shows that the model calculation allows us to 
extract the magnetic cross 
section from the data with some confidence even in this energy range.
The dashed-dotted line describes the prediction of the phonon model which 
accounts well for the 
bowl-shaped background of the scan. The peak in the center of the scan, on 
the other hand, is not 
described by the phonon model and can therefore be recognized as magnetic.

Also, in many cases, additional elements such as the momentum 
dependence of the neutron intensity over several Brillouin zones, its
temperature dependence, its behavior upon changing the resolution 
conditions, etc., have been employed to
cross-check the determination of the magnetic scattering.
Such empirical methods have been 
discussed and successfully applied in previous 
studies \cite{bourges96,bourges97_euro}. 

\section{Results}
\label{sec:results}
Typical constant-energy scans (whose background is generally much more 
benign than that in 
Fig. \ref{fig-sim}) 
are  shown for \YBCO{6.5} in Fig. \ref{fig-typical6.5} and for \YBCO{6.7} 
in Fig. \ref{fig-typical6.7}. Several qualitative features are already apparent from the raw data. 
First, while the magnetic signal is always peaked at or near 
${\bf Q}_0=(\pi,\pi)$, the detailed shape of the profiles 
in momentum space depends on the excitation energy. As the excitation 
energy increases, the peak generally broadens, and at least in the 
\YBCO{6.5} sample it begins to disperse away from ${\bf Q}_0$ above
50 meV (Fig. \ref{fig-typical6.5}). 
The line shape will be discussed in Section \ref{subsec:momentum} below.
Second, the even spin excitations (shown in Fig. 
\ref{fig-typical6.7}) are fully gapped, as are the even excitations 
(optical spin waves) in the 
antiferromagnetic insulator. 
The data were fitted to single or double Gaussians, 
as appropriate, and corrections for the 
resolution function were made. 

The data were further corrected for the 
magnetic form factor of copper. Because of the energy dependence 
of the momentum line shape ( Figs. \ref{fig-typical6.5}-\ref{fig-typical6.7}), we plot 
in Figs. \ref{fig-spectrum6.5} and \ref{fig-spectrum6.7} the 
{\bf q}-averaged (local) susceptibility ($\chi''_{\rm 2D}$, Eq. 
\ref{chi2D}) of the two samples in 
the even and odd channels, derived from the fitted intensities and 
widths of the Gaussian profiles. Alternatively, the 
data can be summarized in terms of the peak intensity at ${\bf Q}_0$,
as shown for \YBCO{6.7} in Fig. \ref{fig-peak6.7}. The local and 
peak susceptibilities are not proportional to each other, because of
the energy dependence of the momentum line shape. Note also that the 
absolute magnitudes of the susceptibilities in 
Figs. \ref{fig-spectrum6.7} and \ref{fig-peak6.7} differ by more 
than order of magnitude, because the susceptibility is always 
strongly peaked at or near ${\bf Q}_0$. 
Nevertheless, the qualitative features of Figs. \ref{fig-spectrum6.5} 
and \ref{fig-peak6.7} are similar. The spin excitations in the 
\YBCO{6.85} crystal are restricted to a much smaller energy range 
than in \YBCO{6.5} and \YBCO{6.7}, as observed previously \cite{revue-cargese}. 
A full spectrum is reported in a
forthcoming publication \cite{bourges00}. 

\subsection{Magnetic Resonance Peak and Superconducting Energy Gap}
\label{subsec:super}
One of the most striking features of the data of Figs. 
\ref{fig-spectrum6.5}a-\ref{fig-peak6.7}a is a pronounced 
peak that develops at low temperatures. Various features of this peak are 
strongly reminiscent of 
the magnetic resonance peak that was observed at an excitation energy of 
40 meV in optimally 
doped \YBCO{6+x} 
\cite{rossat91,mook93,tranquada92,sternlieb94,fong95,bourges96}
and more recently also in $\rm Bi_2 Ca_2 Cu Sr_2 O_{8+\delta}$ 
\cite{fong99_nature}: it is 
concentrated around a single point in an energy-wave vector diagram, its 
intensity decreases with 
increasing temperature, and it occurs in the odd channel. This analogy is 
further bolstered by 
considering the detailed temperature dependence of the peak local 
susceptibility that is shown in 
Fig. \ref{fig-resTDep} for both samples. Clearly, the intensity is sharply 
enhanced below a 
temperature that, to within the experimental error, is identical to $\rm 
T_c$ in both cases. The 
coupling to superconductivity is another characteristic feature of the 
resonance peak in the 
optimally doped compounds. We can therefore clearly associate the 
enhancement of the 
dynamical susceptibility in the underdoped samples below $\rm T_c$ with 
the magnetic 
resonance peak. 

As was noted earlier \cite{dai96,fong97,bourges97_euro}, the energy of the 
resonance peak 
decreases systematically as the hole 
concentration is lowered from optimal doping. Another difference between 
the optimally doped 
and underdoped materials is that in the latter samples magnetic 
excitations are detectable 
in the normal state. The buildup of the magnetic correlations,  
though strongly enhanced in the superconducting state in a narrower 
energy window, already begins much above $\rm T_c$. Indeed, the 
response shown in Figs. \ref{fig-spectrum6.5}-\ref{fig-peak6.7} 
is already broadly peaked in the normal state. 
The normal state spectra are discussed in detail 
in Section \ref{subsec:normal} below.

For now, we focus on the enhancement of the dynamical susceptibility below 
$\rm T_c$. In order 
to elucidate the influence of superconductivity on the magnetic excitation 
spectra, we have 
subtracted the data just above their respective $\rm T_c$ from the data 
deep in the 
superconducting state. The result is shown in Figs. \ref{fig-diff6.5} and 
\ref{fig-diff6.7} 
in the form of $\chi'' ({\bf Q}_0,\omega)$ as well as ${\rm S }({\bf 
Q}_0,\omega)$. In this 
context it is fruitful 
to consider the latter quantity because it is constrained by a sum rule:
\begin{equation}
\frac{\int d^3Q\,d(\hbar\omega)\, 
{\rm Tr}({\rm S}_{\alpha\beta}({\bf Q},\hbar\omega))}{\int 
d^3Q}
=  S(S+1) {\rm\ per\ Cu\ atom}
\label{eq-sumrule}
\end{equation}

This so-called ``total moment sum rule'' is strictly valid only for the 
undoped Heisenberg 
antiferromagnet (with spin $S = \frac{1}{2}$), and the numerical constant 
is likely to be 
somewhat reduced as holes are added in the metallic regime of the phase 
diagram. Independent of 
doping, however, the total moment sum rule implies that ${\rm S}({\bf 
Q},\omega)$, integrated 
over 
all momenta and energies, is temperature independent. Note, however, that 
in a metallic system 
the relevant energies extend up to energies comparable to the Fermi energy, 
far beyond the energy 
range probed by our 
experiment.

Fig. \ref{fig-diff6.5}b shows that for the \YBCO{6.5} sample the 
enhancement of the spectral 
weight around 25 meV, which we attribute to the magnetic resonance peak, 
is accompanied by a 
reduction of spectral weight over a limited energy range both above and 
below 25 meV. The 
width of the response in momentum space does not change significantly upon 
entering the 
superconducting state, and the reduction of spectral weight above and 
below the resonance peak 
compensates the resonant enhancement such that the total moment sum rule 
is satisfied to within 
the experimental error. 

The \YBCO{6.7} sample shows a qualitatively similar behavior, but 
within the energy range probed by our experiment the enhancement of the 
spectral weight around 
the resonance energy of  $\sim$ 33 meV outweighs the reduction of spectral 
weight at other 
energies. This trend continues as the hole concentration in increased. In 
\YBCO{7}, where 
normal state excitations have not been clearly identified, a loss of 
spectral weight accompanying 
the resonance peak has thus far not been observed. At least for doping 
levels exceeding $\rm x 
\sim 0.7$, we therefore have to postulate a broad and weak continuum of 
excitations, perhaps 
extending up to high energies and not directly observed in our 
experiments, as a ``reservoir'' of 
quantum states from which the resonance peak is drawn.

Following these considerations, we have chosen to parameterize the spectral 
weight of the 
resonance peak by the positive component of the difference spectra of 
Figs. \ref{fig-diff6.5} and  
\ref{fig-diff6.7}, $\int d( \hbar \omega) \Delta \chi''_+ ({\bf 
Q}_0,\omega)$, 
in order to compare different doping levels. (Note that in the energy 
range in which 
the enhancement occurs, and over the temperature range considered, 
the Bose population factor $1+n(\omega)$ in Eq. \ref{eq-corrchi} is 
very close to 1, so that $\Delta \chi''_+$ and $\Delta {\rm S}_+$ are 
simply proportional to each 
other.)  Fig. \ref{fig-synopsis} summarizes the doping dependences of the 
spectral weight and 
energy of  the resonance peak as a function of doping concentration. The large error bars
attached to $\Delta \chi''_+$ in Fig. \ref{fig-synopsis} are in part due to
ambiguities arising from the incommensurate response below the resonance peak 
which is discussed below.
While the energy of the resonance peak 
increases with increasing doping level, the absolute spectral weight 
is more weakly doping dependent (though a clear reduction is 
observed in the fully oxygenated sample \YBCO{7}).
Note also that the dynamical susceptibility in the normal state 
decreases strongly with increasing doping level. A feature not shown 
in Fig. \ref{fig-synopsis} is the width of the magnetic resonance 
peak in energy which is comparable to the experimental energy 
resolution for \YBCO{6.7}, \YBCO{6.85} and 
\YBCO{7}, but broadened to an intrinsic width of $\sim$ 10 meV for 
\YBCO{6.5}. It was shown in \cite{fong99} that a small amount
of disorder can lead to a drastic broadening of the peak, and we 
cannot exclude disorder in the Cu-O chain layer as the origin of 
the broadening in YBCO$_{6.5}$.

Although the {\bf q}-width of the resonance peak, $\Delta {\bf q} = 0.25 
{\rm \AA}^{-1}$, does 
not depend significantly on doping, it is nonetheless interesting to 
compute the energy {\it and} 
wave vector integrated spectral weight of the resonance peak, and to 
compare the result to the 
total moment sum rule, Eq. \ref{eq-sumrule}. After a two-dimensional 
Gaussian integration of 
the difference signal, we obtain $0.069 \mu_B^2$, $0.056 \mu_B^2$, 
$0.07 \mu_B^2$, and $0.043 \mu_B^2$ for \YBCO{6.5}, \YBCO{6.7},
\YBCO{6.85}, and \YBCO{7}, respectively. This means that roughly 1-2\% 
of the total magnetic spectral weight is redistributed to the magnetic 
resonance peak upon cooling below $\rm T_c$.

Another important phenomenon at least partly associated with the 
superconducting state
is the formation of a spin gap in the susceptibility spectrum
with decreasing temperature. In the $\rm La_{2-x} Sr_x CuO_4$ 
superconductor, the normal 
state response
is incommensurate and a small spin gap ($\sim$ 3-6 meV) is observed in the 
superconducting state \cite{yamada95,lake99}. In this system, the size of 
the spin gap seems to be
a good indicator of the homogeneity of the samples;
impurities apparently introduce new states below the spin gap, thereby 
smearing out the 
superconductivity-induced anomalies in both temperature and energy. This 
has also been 
confirmed
experimentally in Zn-substituted \YBCO{6+x} \cite{kakurai93,sidis96}.

We have established the size of the spin gap in the superconducting state 
of  \YBCO{6.7} in a 
polarized-beam experiment. Polarization analysis is required for an 
accurate determination of the  
onset of magnetic scattering, as optical phonon scattering is rather 
strong in the vicinity of the 
spin gap.  Fig. \ref{fig-6.7PolTyp} shows typical constant-energy scans 
taken with a polarized 
beam, and Fig. \ref{fig-spingap} shows the low energy spectrum of 
$\rm YBa_2Cu_3O_{6.7}$ derived by fitting the constant-energy scans to 
Gaussians and 
plotting the amplitude versus energy. The large spin gap of $\sim$ 17 meV 
is testimonial to the 
quality of the samples. Its size agrees well with the measurements of 
Rossat-Mignod {\it et al.} on a \YBCO{6.69} sample \cite{rossat93}. 
Similar measurements for \YBCO{6.5} yield an energy gap of about 5 meV at low temperatures.
The doping dependence of the superconducting spin gap is therefore quite different from that of 
the magnetic resonance peak 
itself, as already found previously \cite{rossat93,revue-cargese,revue-lpr}. 

\subsection{Normal State Susceptibility}
\label{subsec:normal}
Having established the influence of superconductivity on the spin 
excitations, we now describe 
their development with temperature and doping in the normal state. 
The general trends can be seen in Figs. \ref{fig-spectrum6.5} and 
\ref{fig-spectrum6.7}, where 
the susceptibility is given in absolute units up to excitation energies of 
120 meV. 
The low-energy spectra in the odd channel, which are consistent with 
previous work \cite{rossat93,revue-cargese,revue-lpr,tranquada92}, 
show a broad peak whose amplitude decreases gradually with increasing 
temperature and doping.  Figs. \ref{fig-spectrum6.5}-\ref{fig-peak6.7} 
demonstrate that the low energy spectral weight 
is already rather small in the normal state, 
and that the additional depression upon entering the superconducting state 
is rather subtle. (This was established for our samples
in detailed studies of the temperature dependence of the low energies, not shown here.)
This is consistent with prior neutron observations of the spin pseudo-gap 
in underdoped \YBCO{6+x}\cite{rossat93} as well as in nuclear 
magnetic resonance (NMR) measurements \cite{yasuoka94,berthier}.
The relationship between the magnetic resonance peak and this 
normal state peak will be described in the next section. 
We here stress that one has to be cautious in interpreting the 
broad normal-state peak in the odd channel simply as a precursor of the magnetic 
resonance peak in the superconducting state. 

Until recently, neutron studies of \YBCO{6+x} have been confined 
to the odd channel, where excitations can be observed at relatively low 
energies. 
This situation was changed when Reznik {\it et al.} \cite{reznik96} 
identified even excitations (optical magnons) in the 
antiferromagnetic insulator. 
Although the coupling between two directly adjacent layers, extracted from 
an analysis of the 
spin wave dispersions, is only $J_\perp \sim 10$ meV, antiferromagnetic 
long range order, 
together with the much larger intralayer exchange coupling of $J \sim 100$ 
meV, 
boosts the gap for optical magnons to $\sim 67$ meV. It is therefore 
interesting to monitor 
the odd-even splitting in the absence of long range order. Figs. \ref{fig-spectrum6.5}b and 
\ref{fig-spectrum6.7}b show that the even spectrum remains fully gapped 
even in the metallic 
regime, although the gap decreases and broadens with increasing doping. 

Similar to the odd channel, the even excitations also exhibit a broad 
peak, albeit centered around 
a higher energy. The temperature evolution of the intensity around the 
peak position is given in 
Fig. \ref{fig-opTDep6.7} for \YBCO{6.7} and compared to the intensity in 
the odd channel at 
the same energy. The even intensity increases markedly with decreasing 
temperature, whereas the 
odd intensity at this energy remains constant to within the error. The 
marked difference in even 
and odd response
functions in this energy range is already quite apparent in the 
data of Fig. \ref{fig-typical6.7}. The gradual 
increase is reminiscent of the gradual growth of the peak intensity in the 
odd channel 
in the normal state (Fig. \ref{fig-resTDep}), but no anomaly is observed 
in the even channel at 
$\rm T_c$. 

We summarize the temperature evolution of even and odd spectra 
schematically in Fig. 
\ref{fig-schematic}.
At high temperatures the dynamical susceptibility is rather featureless. 
Lowering the temperature 
leads to a gradual enhancement of the spin susceptibility in both even and 
odd channels at 
approximately the same rate, but centered around different frequencies. 
Just above $\rm T_c$,  
both spectra therefore look very similar, with a broad peak in each channel 
that is merely shifted in 
frequency. Below $\rm T_c$, the parallel evolution of both spectra ceases. 
The peak in the odd 
channel sharpens abruptly while the intensity in the even channel 
continues its smooth 
normal-state evolution and eventually saturates.

Finally, we draw attention to a noticeable dip in the odd channel
excitation spectrum around 55 meV (Figs. \ref{fig-spectrum6.5} and 
\ref{fig-spectrum6.7}) which we are unable to account for in 
our phonon simulation. Above this minimum the local susceptibility rises to a 
second (less pronounced) maximum. This behavior is in qualitative 
agreement with pulsed neutron data in \YBCO{6.6} \cite{dai99}. 
(Note, however, that the intensity above 70 meV reported in Ref. \cite{dai99} is twice as large as
ours even though both data sets agree in the low energy range). 
The dip is observed in both underdoped samples but is more pronounced 
in \YBCO{6.5}. It is tempting to associate this feature with the gap 
in the even spectrum that also occurs in this energy range. 
A more detailed analysis of the spectra reveals, however, that the gap 
in the even channel has a stronger doping dependence than the 55 
meV feature in the odd channel, which makes this scenario unlikely. 
We briefly discuss alternative explanations in Section 
\ref{sec:discussion} below. Unfortunately, the dip feature occurs 
in an energy region with a nontrivial phonon background 
(Fig. \ref{fig-sim}) and at an energy too high for polarization 
analysis. A detailed experimental characterization is therefore 
difficult.

\subsection{Momentum Line Shape}
\label{subsec:momentum}
In the context of recent debates about the role of  ``charge stripe'' fluctuations in high temperature
superconductivity, the momentum line shape of the spin response in \YBCO{6+x} has emerged as
an important issue. Early indications of an incommensurate response \cite{sternlieb94} were followed up by more 
detailed investigations over a narrow range of energies and doping levels \cite{dai97,mook98}. Because of the 
coarse vertical resolution of a triple axis spectrometer and the nontrivial phonon background, special precautions 
are required to clearly resolve the incommensurate response in \YBCO{6+x}. This is not the primary focus of the 
present study; ongoing detailed measurements of this aspect are focused on the \YBCO{6.85} sample, where the 
incommensurate response is most pronounced, and will be reported separately in a forthcoming publication 
\cite{bourges00}. Here 
we discuss some generic features of the momentum line shape that are apparent without high momentum 
resolution.

\begin{figure}
\caption{
Overall momentum width of the magnetic response in \YBCO{6.7}, obtained as described in the
text (Section \ref{subsec:momentum}). Particular attention was paid to the constant-energy profile
at 24 meV, where an incommensurate response was identified as first observed by Dai et al. \cite{dai97}. Note 
the pronounced broadening above 50 meV. }
\label{fig-momentum}
\end{figure}
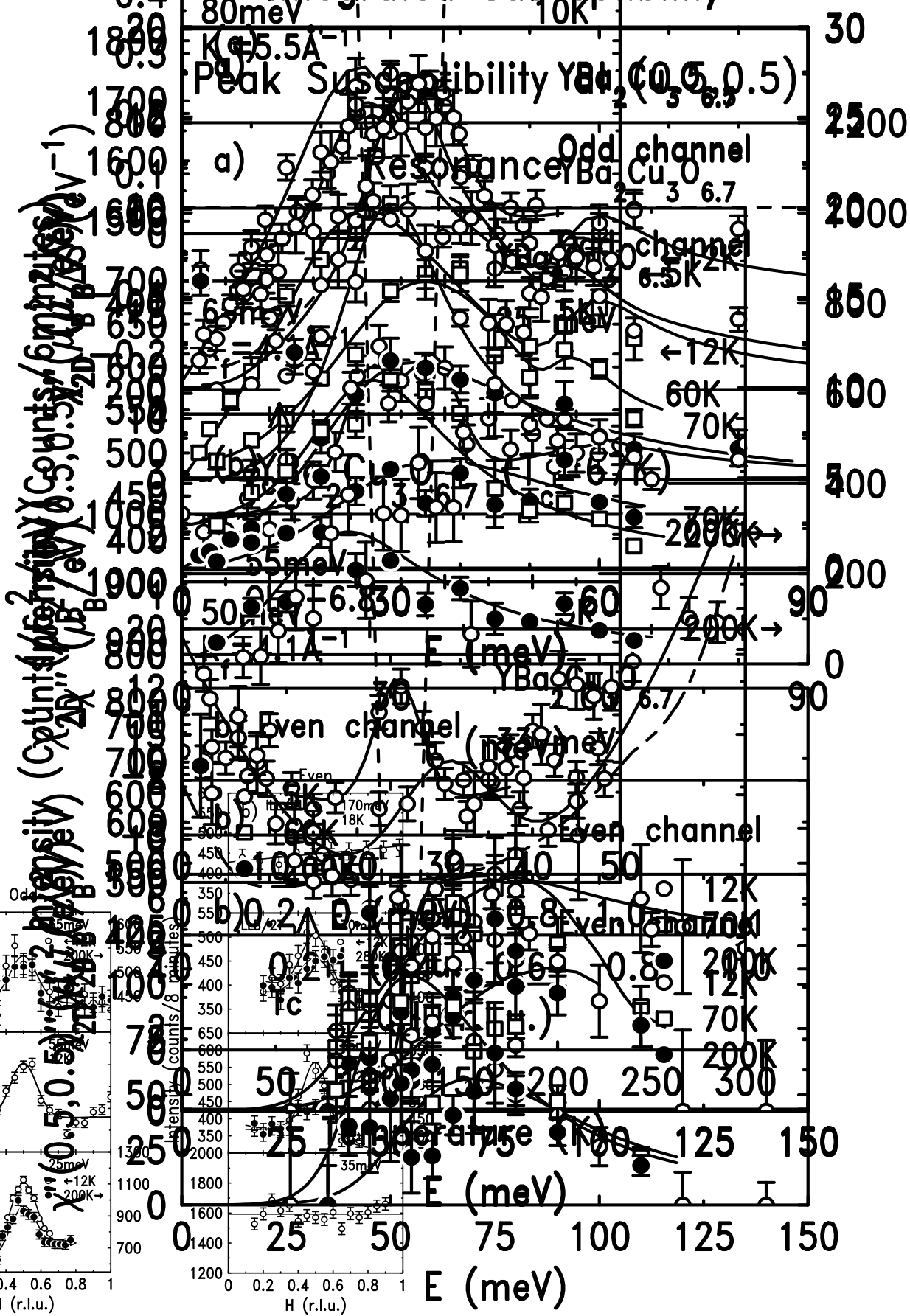
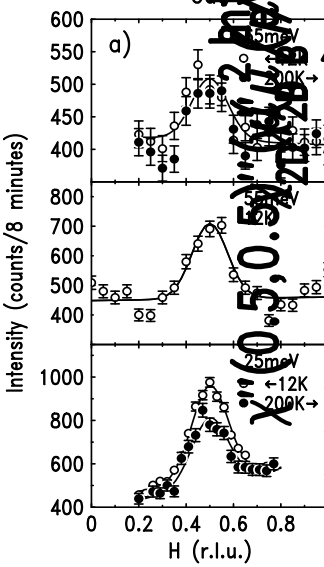
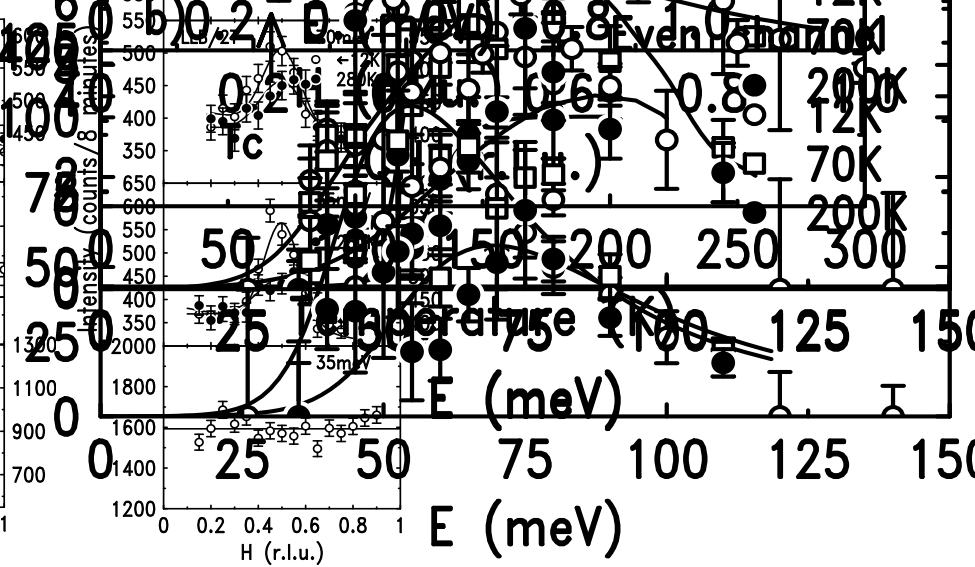
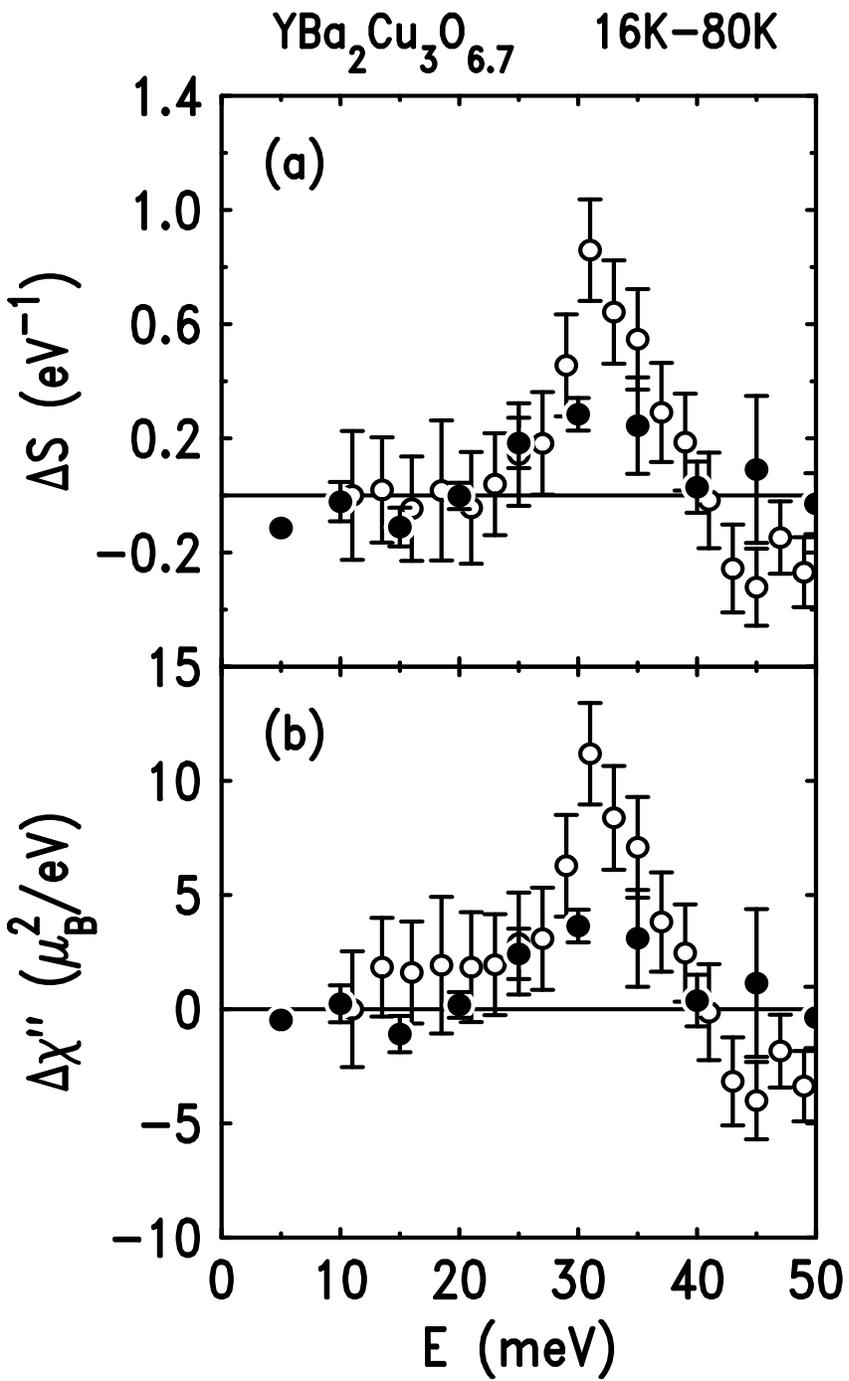
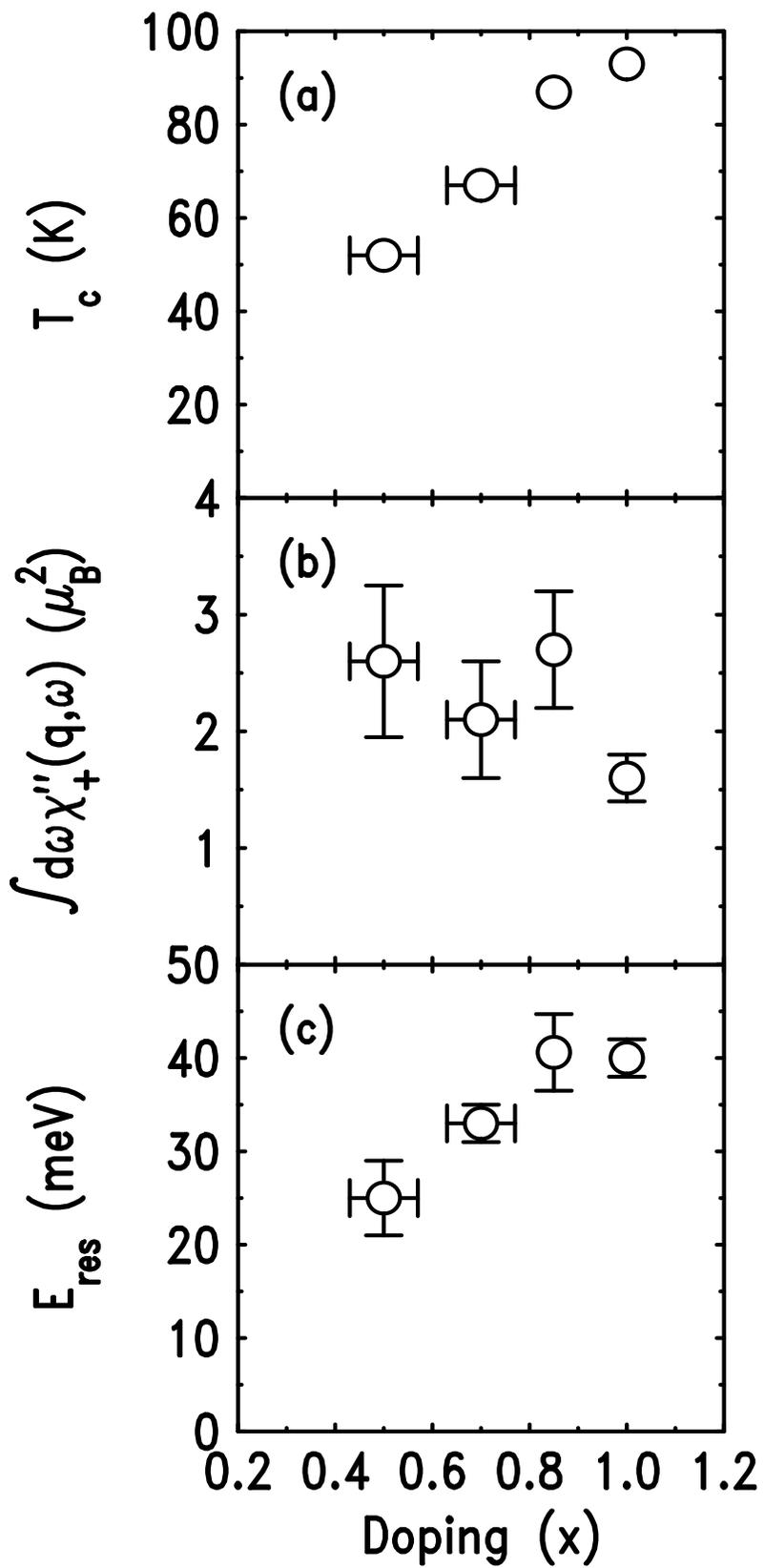
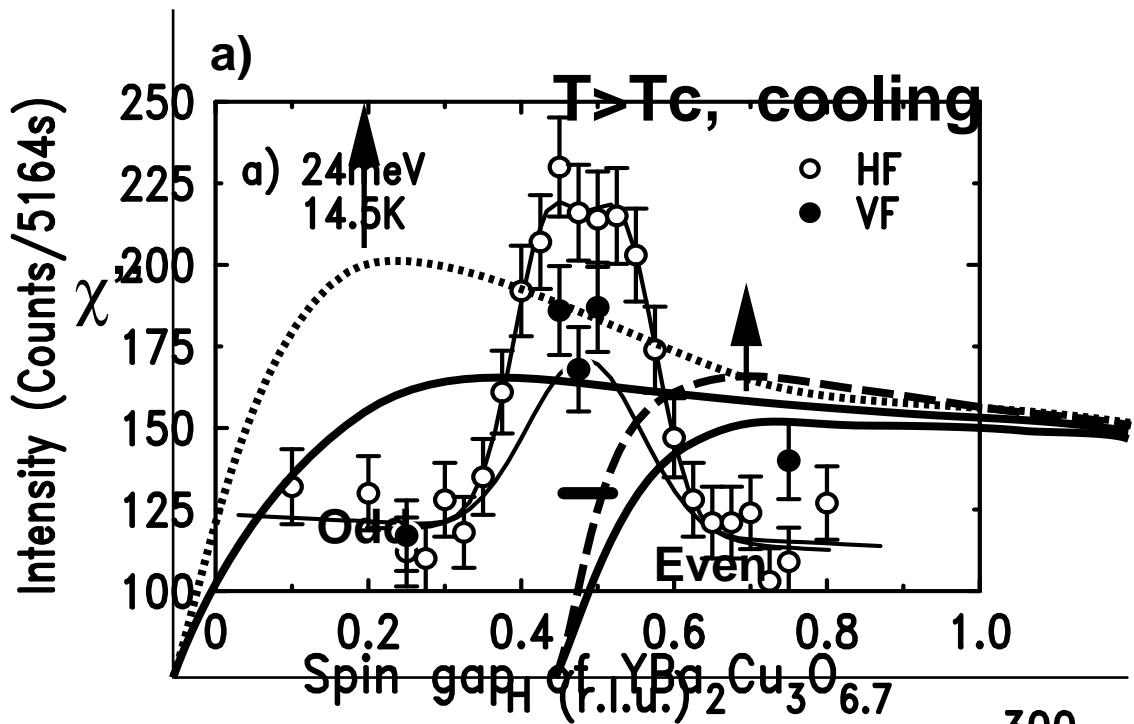
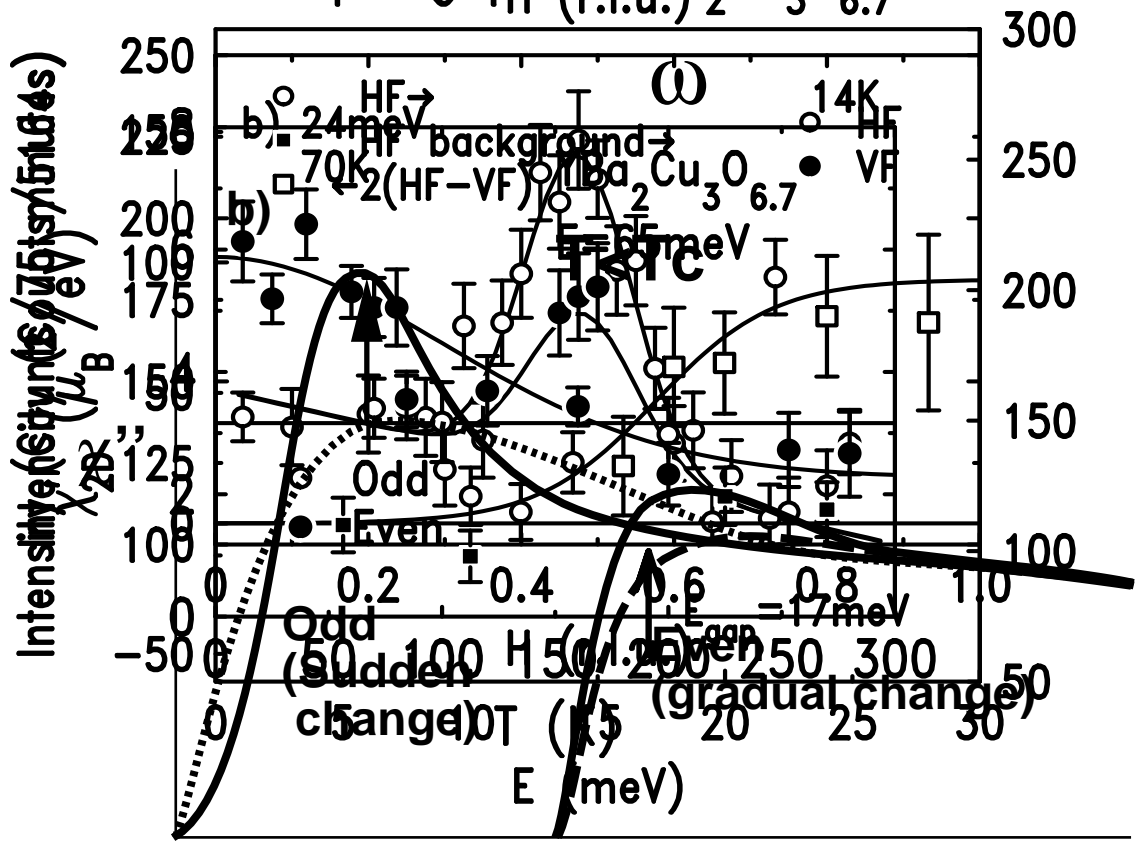
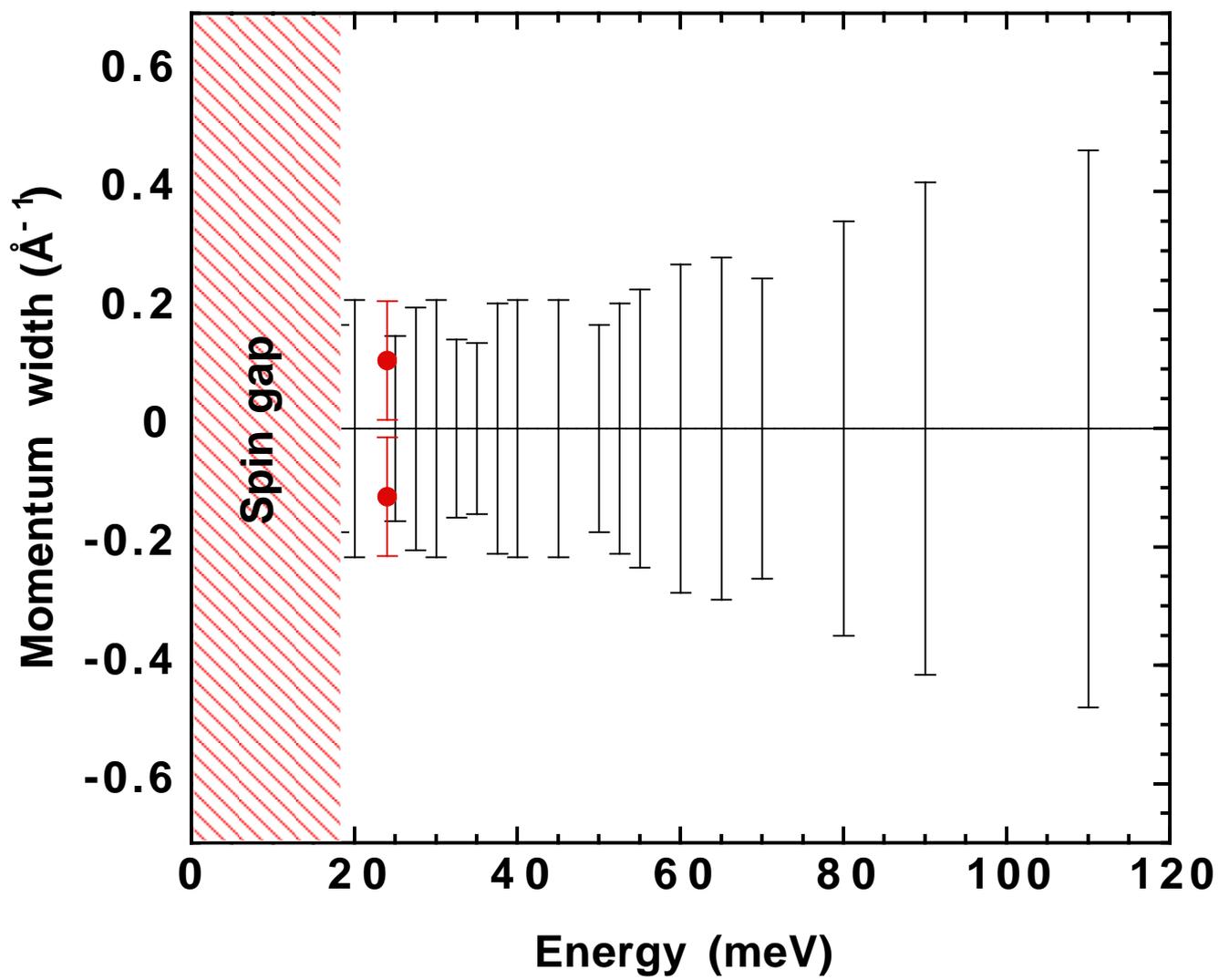

Independent of the detailed line shape, an important parameter for the purposes of the present study is the overall 
extent in momentum space of the magnetic response at a given energy. For instance, this quantity enters into our 
determination of the local susceptibility of Figs. \ref{fig-spectrum6.5} and 
\ref{fig-spectrum6.7}. This overall width was extracted from the constant-energy profiles by fitting them to either 
a single Gaussian centered at  ${\bf 
Q}_0 = (\pi,\pi)$, or if necessary, to two Gaussians symmetrically displaced around  ${\bf 
Q}_0$.  Fig. \ref{fig-momentum} gives the result of this procedure for our \YBCO{6.7} sample. 
While most profiles are adequately fit by a broad single peak, we paid particular attention 
to the energy range around 24 meV, just below the resonance peak , and were able to confirm 
the recent observation \cite{dai97} of an incommensurate response. The weakly energy dependent {\bf q}-width 
below the resonance peak has been noticed before \cite{rossat93,tranquada92,bourges97_euro,balatsky} and may 
reflect an incommensurate response which is unresolved due to the coarse resolution. At the resonance energy, 
the overall {\bf q}-width goes through a minimum which is particularly noticeable when considering the resonant 
response only (that is, the {\bf q}-width of the additional intensity peak below $\rm T_c$, Fig. \ref{fig-diff6.7}) 
\cite{bourges96}. 

The pronounced intrinsic broadening at high energies above 50 meV cannot be attributed to resolution effects and 
is reminiscent of the dispersion-like behavior observed in \YBCO{6.5} (Fig. 
\ref{fig-typical6.5}, see also \cite{bourges97_prb}) which  in turn resembles the dispersion of antiferromagnetic 
magnons. It is interesting to notice that the onset of this broadening coincides with the peak-dip feature in Figs. 
\ref{fig-spectrum6.5} and 
\ref{fig-spectrum6.7}. In this high energy region, the primary effect of hole doping therefore appears to be a 
strong broadening of the magnetic response. Overall, Fig. \ref{fig-momentum} is in good agreement with the 
recent report by Arai {\it et al.} \cite{arai} on a sample with a similar $\rm T_c$. (Note, however, that these 
authors chose to fit all of their constant-energy profiles to double peaks.) 

\section{Discussion}
\label{sec:discussion}
The most striking feature in the neutron spectra  of \YBCO{6+x} is the 
magnetic resonance peak
whose experimental properties are summarized in Fig. \ref{fig-synopsis}. 
This phenomenon has
been addressed in numerous theoretical studies over the past few years. 
One purpose of the 
present
article is to present a summary of the current state of experimental 
information on the resonance
peak. The other purpose is to stimulate further theoretical work on the 
interplay between normal
state excitations and the resonance peak in the underdoped regime. We 
discuss these two 
aspects
in turn.

Compared to the broad continuum expected for ordinary metals, the 
resonance peak involves an 
extraordinarily small volume of phase space. $d$-wave superconductivity 
helps explain this 
observation:
The coherence factor in the 
neutron
scattering cross section has a pronounced maximum at a wave vector 
connecting two lobes of the 
$d$-wave gap function with opposite sign of the order parameter 
\cite{fong95}. (An early 
interpretation
based on a different gap symmetry \cite{mazin95} has been ruled out by a 
variety of other
measurements \cite{van_harlingen95}.) It has since turned out that $d$-wave superconductivity 
alone is not sufficient to 
explain the sharpness and spectral weight of the resonance peak, and that 
other factors further 
narrow the phase space involved in this process. Strong 
Coulomb interactions between electrons are the most likely factor. 

Some of the features of the resonance peak can be described in models that 
do not include such 
interactions. 
For instance, early photoemission studies have suggested an extended saddle 
point singularity near the 
Fermi
level \cite{king94}. It has been shown that this band structure anomaly 
leads to a peak in the 
joint density of 
states, which in turn produces a peak in the unrenormalized susceptibility 
$\chi''_0$ 
\cite{blumberg95,abrikosov98}. 
The peak grows as the gap opens in the superconducting state. The peculiar
gap structure in the interlayer tunneling model of high temperature 
superconductivity also leads 
to a rather
sharp peak in $\chi''_0$  in the superconducting state \cite{lin97}. The 
peak position is 
determined by a 
combination of the energy gap and the chemical potential (see, e.g., Ref. 
\cite{lavagna93}) so that even in this minimalist model the resonance 
peak and the superconducting energy gap do not have to show 
the same doping dependence \cite{onufrieva99}. The resonance peak 
energy {\it decreases} as the hole concentration is lowered from the 
optimally doped state (Fig. \ref{fig-synopsis}) while the energy gap, 
which is directly determined  by photoemission and other techniques, 
remains constant or increases (see, e.g.,  \cite{harris96}). However, the interplay 
between the pseudo-gap measured in the normal state and the 
superconducting gap is not well understood at present in these 
experiments. 

There are, however, strong indications that electron correlations are 
necessary for an adequate 
description
of the resonance peak. First, the absolute magnitude of the dynamical 
susceptibility, determined 
experimentally 
in Refs. \cite{fong96,bourges97_prb}, is too large to be consistent with 
the relatively subtle 
enhancement 
assumed in Refs. \cite{lin97,abrikosov98}. This does not necessarily rule 
out the mechanisms 
proposed 
in this work, but it strongly suggests that Coulomb correlations are also 
needed to reproduce the 
experimentally
measured $\chi''$. Second, the sharpness of the resonance peak close to 
optimal doping indicates 
a true collective mode. This is indeed the direction that most of the 
recent theoretical work has 
taken.

The theoretical description of this collective mode remains a subject of 
considerable controversy. 
In 
one theoretical approach, the peak is identified with a resonance in the 
particle-particle channel. 
Proponents of 
this  model \cite{demler95}, later
embedded into a more comprehensive theory of high temperature 
superconductivity 
\cite{zhang97,demler98_prb}, 
early on predicted the resonance spectral weight to within a factor of two 
of the measured number 
for optimally
doped \YBCO{6+x})  \cite{fong96}. In the framework of this theory, the 
resonance can be 
viewed as a 
pseudo-Goldstone boson of a new symmetry group encompassing 
antiferromagnetism and $d$-
wave 
superconductivity. The symmetry becomes exact at the
quantum critical point separating both phases, and the energy of the 
Goldstone boson approaches 
zero in accord with the data. This scenario is also consistent with the tendency 
of the resonance spectral weight to increase with decreasing hole 
concentration (Fig. \ref{fig-synopsis}). 
All of these features are, however, also consistent with a more 
conventional description in which the resonance
occurs in the particle-hole channel. In this approach, the resonance 
is identified with
a spin exciton that is 
stabilized when decay channels are removed in the superconducting state. Recent 
theoretical studies have shown 
that the evolution of the resonance energy and spectral weight with doping 
(Fig.  \ref{fig-synopsis}) as well as the
broadening of the resonance in the deeply underdoped regime can be 
explained in this framework
\cite{onufrieva99,brinckmann99}. 
It has also emerged that a remarkably consistent picture of ARPES and neutron
data can be obtained in this framework \cite{norman,chubukov,onufrieva99}. An interesting suggestion put forth 
in this context is to interpret
the weakly doping dependent dip-peak feature of Figs. \ref{fig-spectrum6.5} and 
\ref{fig-spectrum6.7}as a signature of the 
bare superconducting energy gap, while the resonance peak is pulled below the gap
by interactions \cite{chubukov}. Further work is necessary to establish whether the onset of the momentum-space 
broadening
in this energy range (Section \ref{subsec:momentum}) can also be explained in this model.
Finally, it has been proposed that the resonance peak originates from localized electrons in domain walls between 
charge stripes, separate from those that
form the superconducting condensate \cite{emery97}. While the large spectral weight of the peak finds a natural
explanation in this approach, it is much harder to account for its sharpness in both energy and momentum and its
strong coupling to superconductivity. Further, the evidence of charge stripes is weak in optimally doped 
\YBCO{7} where the resonance peak is most pronounced.

We do not wish to reiterate the purely theoretical arguments that were 
advanced in this debate 
(see, e.g., 
\cite{demler98_prb,greiter97,baskaran98,brinckmann99,onufrieva99}). 
Rather, we point
to the additional experimental information presented in this article, much 
of which still awaits a
theoretical explanation. We first discuss the normal state susceptibility 
described in Section 
\ref{subsec:normal}
and summarized schematically in Fig. \ref{fig-schematic}. The broad peak 
in the odd channel 
that grows with
decreasing temperature (already been observed in early work 
on underdoped  \YBCO{6+x} 
\cite{rossat93,revue-cargese,revue-lpr,tranquada92}) signifies a new energy scale in the normal-state spin 
excitation spectrum, different from the superexchange interaction $J$ that sets the energy scale in the insulator. It 
was 
pointed out \cite{barzykin95,tranquada97} that damping of 
antiferromagnetic spin waves (seen as long-lived excitations in 
the insulator) introduces such a new energy scale, namely $\sim J a/\xi$ (where $\xi \sim 2a-5a$, depending on 
doping,
is the spin-spin correlation length), above which the excitations are not substantially different from those of the 
insulator. This approach can describe some aspects of the data, especially the dispersion-like behavior observed at 
high energies in \YBCO{6.5}. The response in 
the even channel is 
also qualitatively consistent with a description based on damped 
spin waves. When $\xi$ is short, one may expect the even and odd 
response functions to be shifted by $J_\perp \xi/a$, the energy cost for flipping a correlated
patch of spins in a single layer while keeping the spins in the neighboring layer fixed. (This
simple estimate was confirmed by numerical simulations on finite-sized systems.) The 
actual situation (Figs. \ref{fig-spectrum6.5} and \ref{fig-spectrum6.7}) 
interpolates between this limit and the antiferromagnetic long range 
ordered state with an optical magnon gap 
of $2 \sqrt{J J_\perp} \sim 67$ meV \cite{reznik96}. 

The temperature dependence of the broad normal-state peak in the pseudo-gap regime and its relation to the 
magnetic resonance peak are 
more difficult to describe.
Of course, short range dynamic spin correlations centered around ${\bf 
Q}_0 = (\pi,\pi)$ are 
expected in a variety 
of  microscopic strong-correlation models. The challenge is 
now to use these models to make detailed predictions of the temperature 
evolution of the 
magnetic spectra in both 
the superconducting and the normal states.  Phenomenologically, the low-energy
``overdamped spin wave'' 
response
grows with decreasing temperature. The broad peaks in both even and odd 
channels 
grow gradually and approximately at the same rate, while at the same time 
a spin pseudo-gap 
opens in the odd 
channel as already revealed by earlier neutron \cite{tranquada92,rossat93} 
and NMR \cite{yasuoka94} experiments.
Although it is hard to associate a characteristic temperature with this 
crossover phenomenon, our 
data are
certainly consistent with models that predict a correlated spin liquid 
phase below a doping
dependent ``coherence temperature'', often termed $\rm T^*$ 
\cite{fukuyama96}. 

It is important to draw a distinction between the gradual growth of 
the broad normal state 
peaks in both
even and odd channels on the one hand, and the
abrupt increase of the intensity in the superconducting state that occurs 
in the odd channel {\it 
only} on the
other hand. This is seen most clearly in the high-statistics data of Fig. 
\ref{fig-resTDep}a. Though 
both are clearly 
related, a 
recent  re-interpretation of the previously observed broad normal-state 
peak in the odd channel 
as a ``pre-transitional precursor'' of the resonance peak  is therefore 
too simple-minded 
\cite{dai99}. There is, 
moreover, no basis for a 
separation of 
resonant and non-resonant response functions in the normal state as 
proposed in Ref. 
\cite{dai99}. Promising 
steps in the 
direction of a microscopic description of the spin dynamics in both normal 
and superconducting 
states have 
recently been  taken (see, e.g.,  Ref.  \cite{weng98}).

This also bears directly on the issue of the origin of the superconducting 
condensation energy that 
has recently 
moved into the foreground of the high-$\rm T_c$ debate. Scalapino and 
White 
\cite{scalapino98} derived a 
formula that relates the exchange energy (associated with the $J$-term in 
the $t-J$ model) to the 
magnetic 
excitation spectrum. If the complete spectrum is known in absolute units, 
the change in exchange 
energy between normal 
and superconducting states can be evaluated and quantitatively compared to 
the superconducting 
condensation 
energy determined in specific heat experiments \cite{loram}. In optimally doped 
\YBCO{6+x}, the resonance 
peak is the only experimentally discernible feature in the magnetic 
spectrum, and its spectral 
weight is known in absolute units \cite{fong96}. Since the normal state 
spectrum is unknown 
except for an upper bound guaranteeing that the amplitude of the normal 
state excitations is 
significantly below the resonance amplitude \cite{bourges96,fong96,miami},
it is reasonable to apply the Scalapino-White formula to the resonance 
peak only. This is the approach adopted by
Demler and Zhang \cite{demler98_nature} who found that the magnetic energy 
stored in the 
resonance peak equals the superconducting condensation energy to within an 
order of magnitude. 
Temperature 
dependent changes in other parts of the spectrum, though not visible in 
the neutron experiments, could of course lead to quantitative modifications in this analysis. The full implications 
of this analysis for the pairing 
mechanism of high 
temperature superconductivity are, however, still under debate. Others \cite{kee99} have argued,
for example, that the neutron peak provides a measure of the condensate fraction, rather than the
condensation energy, of the superconducting state.

The experimental situation is much less straightforward in the underdoped 
regime. We have shown here that the amplitude of the magnetic response 
in the normal state is at least as large as the resonance amplitude, 
and that it is spread over a much larger energy range. 
If one wants to extend the Demler/Zhang analysis into the normal state 
of underdoped \YBCO{6+x}, it is therefore erroneous to 
focus exclusively on the energy at which the resonance 
eventually develops in the superconducting state \cite{dai99}, especially since a separation of the normal-state 
response into resonant and nonresonant contributions is entirely arbitrary. It is also 
inappropriate to neglect the 
optical channel in 
such an analysis, because its amplitude and temperature dependence in the 
normal state are at 
least equal to those 
in the superconducting state. Because the Scalapino-White formula differs from the total moment sum rule (Eq. 
\ref{eq-sumrule}) only through a momentum-dependent form factor, any gain in magnetic energy must result from 
temperature dependent changes of the momentum line shape which are not even considered in Ref. \cite{dai99}. 
It is easy to see that the outcome of an adequate analysis of 
the magnetic energy 
stored in the 
normal state spin excitations and its relation to the specific heat would 
likely be qualitatively different from the one given in Ref. 
\cite{dai99}.  In particular, the normal-state electronic specific heat in the pseudogap regime of the \YBCO{6+x} 
phase diagram is actually {\it lower} than at optimum doping \cite{loram}, whereas the magnetic spectral weight 
and its temperature dependence are clearly {\it larger} in underdoped samples.
Because of insufficiently accurate information on the temperature dependence of the momentum line shapes, we 
have not used our spectra to attempt 
such an analysis ourselves. 

In summary, we have presented a comprehensive account of the magnetic resonance peak and the 
normal state spin excitations in underdoped \YBCO{6+x}. We hope that the detailed description of the magnetic
spectra presented here will provide an improved basis for models of electronic correlations in the cuprates.
\acknowledgements
We wish to thank M. Braden at 
LLB and J. Kulda at ILL for their 
help during the experiments, and P.W. Anderson, A. Chubukov, V.J. Emery, B. Hennion, M. Norman, F. 
Onufrieva, and P. 
Pfeuty
for stimulating discussions.

\begin{appendix}
\section{Phonon simulation}
\label{apdx-sim}
Empirical descriptions of the eigenvibrations in a single crystal are
instrumental in obtaining the absolute magnitude of the dynamical
susceptibility in our measurements. However, as there are 39 phonon
branches in \YBCO{7} spread over an energy range of roughly 70 meV, individual
phonon branches are usually difficult to resolve (with a 
typical energy resolution of 5 
meV). Fortunately, in past
successful experiments, we have isolated one particularly strong phonon
branch at 42.5 meV \cite{fong95,fong96}. 
Since then we have further reinforced confidence in
our model by identifying and simulating acoustic phonon branches. 
Acoustic phonons are attractive for absolute unit measurements
because their intensities are strongly related to the Bragg peak 
intensities.
Unlike optical phonons, the most important quantity for the acoustic 
phonons
is just one number-the limiting value of the structure 
factor as it approaches the Bragg condition.
The convolution  of their ``cone shape'' dispersion curve 
with  the instrumental resolution  has been performed 
analytically as well as by using an efficient sampling technique 
based on the fast
Fourier transform to perform a four dimensional numerical convolution
\cite{fftw}. We have extensively used a longitudinal acoustic 
phonon branch around
the (006) Bragg condition. Our simulation gives a structure factor of
$4.3\times 10^{-18} c^2\,{\rm eV}^{-1}$ which is consistent with our 
earlier
optical phonon calibration measurements. We further point out that 
the calibration using phonon scattering from the sample avoids 
the difficulties related to existence of ``green phase'' 
impurities as well as inclusions of smaller crystallites inevitable 
in these large \YBCO{6+x} samples. This may contribute to 
systematic errors when calibrating against a 
vanadium standard \cite{hayden96,dai99}.

\end{appendix}

\begin{figure}
\caption{
$T_c$ curve of the \YBCO{6.7} sample using the depolarization technique 
as described in the text.  The spectrometer is set for the (006) nuclear 
Bragg reflection. The 
flipping
ratio, defined as the ratio of neutrons scattered without spin flip to 
those that are scattered with a 
spin flip, 
is infinite at a nuclear Bragg reflection in an ideal setup, but finite 
(here: 16) because of 
incomplete beam 
polarization. With the sample field cooled into the superconducting state 
and then rotated by 
90$^\circ$, 
the flipping ratio is reduced by trapped flux.
}
\label{fig-6.7Tc}
\end{figure}

\begin{figure}
\caption{
A scan of even channel excitations at 55 meV that exemplifies the 
usefulness
of numerical simulations of the lattice dynamics. The dotted line, 
derived from the phonon simulation program described in the text, 
provides a good description of the background features. The peak in the 
center is magnetic.
The solid line is a guide-to-the-eye. 
}
\label{fig-sim}
\end{figure}

\begin{figure}
\caption{
Typical constant-energy scans taken on the $\rm YBa_2 Cu_3 O_{6.5}$ sample 
in the odd 
channel. The solid 
lines are the results of fits to Gaussian profiles. The energies
of the data in the three panels are evenly spaced (50 meV, 65 meV, and 80 
meV). The dotted 
lines that join the
split peaks  go through  $(H = 0.5 {\rm r.l.u.},\ E= 0{\rm meV})$ and 
illustrate a spin-wave-like dispersion, 
as described in the text.
}
\label{fig-typical6.5}
\end{figure}

\begin{figure}
\caption{
Typical constant-energy scans obtained on the $\rm YBa_2 Cu_3 O_{6.7}$ 
sample, in
(a) odd and (b) even excitation channels. The top panel of  (b) shows ILL 
data, the
rest are from LLB experiments. The lines are the results of fits to 
Gaussian profiles.
}
\label{fig-typical6.7}
\end{figure}

\begin{figure}
\caption{
Local (2D wave vector averaged) susceptibility of $\rm YBa_2 Cu_3 O_{6.5}$
in the (a) odd and (b) even excitation channels, obtained by fitting 
constant-energy scans..
Solid lines are guides-to-the-eye. The dotted line indicates the low-
energy local susceptibility of
the 2D Heisenberg antiferromagnet (Eq. \ref{chiAF}).
}
\label{fig-spectrum6.5}
\end{figure}

\begin{figure}
\caption{
Local (2D wave vector averaged) susceptibility of $\rm YBa_2 Cu_3 O_{6.7}$
in the (a) odd and (b) even excitation channels. Solid lines are guides-
to-the-eye. The dashed line 
superposed on 
the top (12 K) spectrum reproduces the solid line for the 70K data and 
illustrates the effect of 
entering the 
superconducting state.  The dotted line indicates the low-energy local 
susceptibility of
the 2D Heisenberg antiferromagnet (Eq. \ref{chiAF}). 
(from \cite{bourges97_prb}).
}
\label{fig-spectrum6.7}
\end{figure}

\begin{figure}
\caption{ Same data as in Fig. \ref{fig-spectrum6.7}, but showing the intensity at ${\bf 
Q}_0=(\pi,\pi)$ instead of 
the local susceptibility.
}
\label{fig-peak6.7}
\end{figure}
 
\begin{figure}
\caption{
Temperature dependence of the local susceptibility at the energies at 
which the magnetic 
resonance occurs in both 
underdoped samples. Superconducting critical temperatures are marked with 
arrows. The solid 
symbols are 
obtained using polarized neutron scattering techniques.
}
\label{fig-resTDep}
\end{figure}

\begin{figure}
\caption{
Difference of (a) the susceptibility $\chi'' ({\bf Q}_0,\omega)$ and (b) 
the spin-spin 
correlation function,  ${\rm S }({\bf Q}_0,\omega)$ in the superconducting 
and normal states, 
for 
\YBCO{6.5}. Normal state data are taken at 60 K, just above $\rm T_c = 52$ 
K. 
}
\label{fig-diff6.5}
\end{figure}

\begin{figure}
\caption{
Difference of (a) the susceptibility $\chi'' ({\bf Q}_0,\omega)$ and (b) 
the spin-spin 
correlation function,  ${\rm S }({\bf Q}_0,\omega)$ in the superconducting 
and normal states, 
for 
\YBCO{6.7}. Normal state data are taken at 80 K, just above $\rm T_c = 67$ K. The open 
circles represent data taken at BNL, the filled circles are data taken at LLB. }
\label{fig-diff6.7}
\end{figure}

\begin{figure}
\caption{
A synopsis of (a) the superconducting transition temperature, (b) the energy-integrated spectral weight evaluated 
at  ${\bf Q}_0=(\pi,\pi)$, and 
(c) of the energy 
of the magnetic resonance peak in the two underdoped samples, compared to 
those of the 
optimally doped sample with $\rm T_c=93$ K. Horizontal error bars indicate 
a conservative 
confidence level for 
the oxygen content.
}
\label{fig-synopsis}
\end{figure}

\begin{figure}
\caption{
Typical constant-energy scan at 24 meV taken with polarized neutrons in the 
spin-flip channel on 
$\rm YBa_2 
Cu_3 O_{6.7}$ in (a) the superconducting and (b) the normal state. Open 
circles are horizontal 
field (HF) 
measurements taken after the sample was field cooled with the field 
parallel to the scattering 
vector. Solid circles are measurements taken after the sample was cooled in 
vertical field (VF), 
perpendicular to the 
scattering plane. For magnetic scattering, the background-corrected HF 
intensity should be twice 
the VF intensity 
because of polarization factors in the magnetic neutron cross section 
(Eq. \ref{hfvf}). The 
solid lines indicate that 
this is indeed 
the case.
}
\label{fig-6.7PolTyp}
\end{figure}

\begin{figure}
\caption{
Determination of the spin gap in $\rm YBa_2 Cu_3 O_{6.7}$ by polarized 
neutrons. The lines are guides-to-the-eye. Open square symbols are from direct
subtraction between HF and VF intensity as in
the previous figure. Solid square symbols are HF
polarized beam background from curve fitting such data. Open circles are
the amplitudes resulting from these fits. Note that the scales on left and right 
vertical
axis are the same.
}
\label{fig-spingap}
\end{figure}

\begin{figure}                                                              
\caption{
Local susceptibility at 65 meV in $\rm YBa_2 Cu_3 O_{6.7}$ in
both channels, showing their different temperature dependences.
Note the absence of any anomaly at $\rm T_c$. Lines are guides-to-the-eye.
}
\label{fig-opTDep6.7}
\end{figure}

\begin{figure}
\caption{
Schematic diagram summarizing the temperature evolution of the
magnetic excitation spectra of the underdoped \YBCO{6+x} compounds (a) in 
the normal state, and (b) across the superconducting transition.
The even excitations evolve smoothly with temperature, parallel to (a).
The odd channel excitations undergo an abrupt sharpening at
the resonance energy across $\rm T_c$.
}
\label{fig-schematic}
\end{figure}

\end{document}